\documentclass[10pt, onecolumn]{IEEEtran}
\IEEEoverridecommandlockouts
\usepackage{soul}
\usepackage{amsmath}
\usepackage{algorithm}
\usepackage{algpseudocode}
\usepackage{array}
\usepackage{subcaption}
\usepackage{amsmath,amssymb,amsfonts}
\usepackage{algorithm,algpseudocode}
\usepackage{graphicx}
\usepackage{textcomp}
\usepackage{xcolor}
\usepackage{romannum}
\usepackage{multirow}
\usepackage{placeins}
\usepackage{mathrsfs}
\usepackage{ulem}
\usepackage{url}

\def\BibTeX{{\rm B\kern-.05em{\sc i\kern-.025em b}\kern-.08em
    T\kern-.1667em\lower.7ex\hbox{E}\kern-.125emX}}
\begin{document}
\title{Deterministic Scheduling over Wi-Fi 6 using Target Wake Time: An Experimental Approach}%
\author{\IEEEauthorblockN{Govind Rajendran and Samar Agnihotri}\\%
Correspondence Email: govindrajendran98@gmail.com and samar.agnihotri@gmail.com
}

\maketitle

\begin{abstract}
\label{subsec: abstract}
Wi-Fi networks traditionally use Distributed Coordination Function (DCF) that employs CSMA/CA along with the binary backoff mechanism for channel access. This causes unavoidable contention overheads and does not provide performance guarantees. In this work, we outline some issues that occur with the probabilistic channel access in highly congested scenarios and how those can be mitigated using deterministic scheduling. Towards this, we propose to use Target Wake Time (TWT) - a feature introduced in Wi-Fi 6 as a power-saving mechanism, to improve the performance of Wi-Fi. To gain insights into the workings of the TWT over commercially available off-the-shelf components and to analyze the factors that affect its performance, we carry out various experiments with it over our Wi-Fi 6 testbed. Using these insights and analysis, we formulate and solve an optimization problem to synthesize deterministic schedules and obtain the optimal values of various system parameters. Lastly, we configure our testbed with these optimal parameter values and show that the TWT based deterministic scheduling consistently results in better performance of the TWT-capable clients and overall system performance compared to traditional CSMA/CA based scheduling.
\end{abstract}


\section{Introduction}
Wireless communication technologies have played an important role in the current information age and Wi-Fi is an integral part of this revolution. Allowing easy deployment and better indoor performance makes it the de-facto method to connect consumer devices to the Internet nowadays, replacing traditional wired connection. It is estimated that approximately 200 exabytes of global monthly IP traffic is transmitted through Wi-Fi \cite{b1}. Along with the major role it plays in smart devices, Internet of Things (IoT) infrastructure and enterprise deployments have made Wi-Fi a common utility. Wireless Local Area Network (WLAN), which is based on the IEEE 802.11 suite of standards and is commonly known as Wi-Fi, is a popular choice for network connectivity. However, its popularity comes at the cost of efficiency and performance as Wi-Fi devices share an unlicensed and unregulated band. As Wi-Fi, traditionally, uses contention as its primary channel access method, it can incur a significant penalty in terms of throughput and delay, affecting the Quality of Service (QoS) experienced by the users. Airtime is a scarce resource, more so in a congested scenario, and the need for its optimal utilization to provide QoS guarantees to the end-users and to improve the overall system performance cannot be overstated.

A newer generation of Wi-Fi, known as High Efficiency Wireless or Wi-Fi 6, built according to the specifications of the IEEE 802.11ax standard \cite{b20}, focuses on more efficient spectrum utilization and providing QoS rather than higher throughput. Wi-Fi 6 introduces 1024-QAM, Orthogonal Frequency Division Multiple Access (OFDMA), 8-stream multi-user multiple input multiple output (MU-MIMO), and spatial reuse among other features to improve the end-user performance and increase spectral efficiency even in dense deployment scenarios. Additionally, it also has \textbf{Target Wake Time}, a feature initially introduced in 802.11ah standard \cite{b18} to reduce signal overhead and network energy consumption.

Carrier-Sense Multiple Access with Collision Avoidance (CSMA/CA) is a channel access method traditionally used by Wi-Fi devices. A CSMA/CA client listens to check for an idle channel before transmission. If the channel is indeed busy, it waits till the end of the transmission plus the Distributed Inter Frame Space (DIFS) duration before choosing a random delay. This delay is uniformly chosen from the contention window. If any other client transmits during this period, the countdown is paused and resumed after the transmission is completed. In case there is a collision, the client uses a binary exponential back-off mechanism before trying again, wherein the contention window is doubled, and a new delay is sampled from it. The back-off, on average, doubles after every failure, thus reducing the collision probability.

However, the CSMA/CA based channel access may cause significant penalties in terms of delay and throughput, while reducing the overall efficiency of the system due to airtime wasted because of the back-off mechanism. The CSMA/CA also impacts the fairness of the transmission. In a scenario where the traffic is downlink biased, yet a few clients with uplink traffic may receive a higher proportion of transmit time. This is because over a period of time the odds of every device winning the channel average out to be the same and when the Access Point (AP) wins the channel, its transmit time must be further divided among multiple downlink clients. This is often referred to as the uplink/downlink disparity of probabilistic channel access \cite{b21}.

The CSMA/CA due to its inherent randomness in providing the channel access, cannot provide guarantees on airtime, throughput, or delay jitter, and it suffers significantly as the number of associated clients increases. Another issue with the probabilistic access arises with the hidden terminal problem \cite{b19}, when clients are within range of the AP, but out of range with each other. It is tackled with the RTS/CTS mechanism which adds to the transmission overhead.

\subsection{Deterministic channel access}
\label{subsec: Determisinitic Scheduling}
Scheduling packets to be transmitted at predefined time slots, without contending for channel access, unlike CSMA/CA, is defined as deterministic channel access. This allows us to give guarantees on channel access, which may lead to guarantees on throughput and jitter while also taking into consideration the QoS demands and fairness.

In commercial enterprise deployments like office and residential spaces, channel usage patterns of applications and clients can be captured and analyzed, along with fine-grained information such MCS index, signal strength, to build spatio-temporal models of usage. This can be leveraged effectively to set up different operational Wi-Fi parameters/schedules to handle the expected demands~\cite{b24}. While mobility of clients is a practical feature of wireless networks, scheduling can still allow some guarantees/increase in performance, for example in scenarios such as legacy channel access and limited mobility.

We propose to use the Target Wake Time feature in Wi-Fi 6 to provide a predetermined channel access schedule, allowing transmission of packets at specific predefined intervals and show its effectiveness over legacy CSMA/CA based channel access. To the best of our knowledge, our work is the first demonstration of this idea on an experimental WLAN testbed, composed of only consumer-grade off-the-shelf components.

\subsection{Contributions}
\label{subsec: Contributions}
In this work, we
\begin{itemize}
\item study the throughput characteristics of Wi-Fi traffic with the variation of airtime through a series of experiments performed on our Wi-Fi 6 testbed designed with commercially available off-the-shelf TWT-capable Wi-Fi 6 clients and other components,
\item propose an optimization problem to create deterministic schedules in a WLAN with TWT-capable clients and to optimize various system parameters to maximize the throughput of such a system, subject to the constraints of fairness and minimum throughput guarantees,
\item deploy the deterministic schedules and optimum values of various system parameters in our Wi-Fi 6 testbed, and evaluate the resulting performance and compare it with a system with only legacy CSMA/CA clients,
\item and experimentally show that in a hybrid environment with a mix of TWT-capable and traditional clients, the proposed approach not only results in higher throughput for the individual TWT-capable clients, but also consistently provides higher overall system throughput.
\end{itemize}

\subsection{Organization}
\label{subsec: Organization}
The paper is organized as follows. In Section~\ref{sec: TWT Section}, we introduce TWT. Section~\ref{sec: Related Work} provides an overview of related work. Sections~\ref{sec: Experimental setup} and \ref{sec: Performance Analysis of TWT} describe, respectively, the experimental testbed and performance analysis of TWT over it to gain insights into working of TWT available on off-the-shelf components. Based on these insights, in Section~\ref{sec: Optimization Problem}, we formulate an optimization problem to find the optimal TWT-based deterministic schedule and system parameter values to maximize the overall throughput, and in Section~\ref{sec: Solution approach}, we offer a practical approach to solve this problem. Using the optimal values of various system parameters thus obtained for our WLAN testbed, in Section~\ref{sec: Results} we configure the testbed accordingly and experimentally compare the performance of the system with the proposed deterministic scheduling and the legacy random channel access. Our results show that deterministic scheduling not only improves the performance of individual TWT clients, but of the overall system as well. Finally, Section~\ref{sec: Conclusions} concludes the paper and provides some directions for the future work.

\section{Target Wake time}
\label{sec: TWT Section}
Target Wake Time (TWT) is a feature that was first introduced in the IEEE 802.11ah \cite{b18} standard as a power saving mechanism for devices, such as battery-limited IoT devices, that need to transmit data periodically, yet have a critical need for long lifetimes. TWT is a periodic wake and sleep pattern negotiated between the AP and the clients \cite{b7} allowing the clients to wake up during the predefined intervals. This helps in reducing the power consumption of the devices while also reducing the overall contention for the channel. This feature is further upgraded in 802.11ax, allowing triggered transmissions and Broadcast TWT to improve the multi-user capabilities \cite{b3}.

\subsection{Types of TWT}
\label{subsec: Types of TWT}
\subsubsection{Individual TWT}
\label{subsubsec: Individual TWT}
In this mode of operation, a client and the AP negotiate wake and sleep periods, which allow the AP to exactly know when the client is available for transmission/reception. It is important to note that when there are multiple clients waking up at the same time, the CSMA/CA channel contention rules apply.

\subsubsection{Broadcast TWT}
\label{subsubsec: Broadcast TWT}
In this mode of operation, a common wake period is shared among a group of clients, with Broadcast TWT specific negotiation parameters similar to individual TWT. Further, clients of this shared session may opt to use OFDMA to optimize their transmissions. It is to be noted that most commercially available devices do not yet support Broadcast TWT.

\subsection{Individual TWT parameters}
\label{subsec: TWT parameters}
Since this work primarily deals with individual TWT mode and its applications, next we discuss some of its major operational parameters with reference to Figure~\ref{fig:TWT-Illustration}.

\subsubsection{Target Wake Duration}
\label{subsubsec: Target Wake duration}
It refers to the minimum duration for which a client remains awake for transmitting or receiving data (the spikes in Figure~\ref{fig:TWT-Illustration}). We refer to this as the \textbf{waketime (WT)} for the reminder of the paper.

\subsubsection{Target Wake Interval}
\label{subsubsec: Target Wake Interval}
It is the time between two consecutive WTs during which a client may choose to sleep (the valleys between the WTs in Figure~\ref{fig:TWT-Illustration}). We refer to this as \textbf{sleeptime (ST)} for the reminder of the paper. It is observed that if a client has traffic to send, it may choose not to sleep and instead continue to transmit during the ST.

\subsubsection{Target Wake Time}
\label{subsubsec: Target Wake Time}
\begin{figure}[!t]
    \centering
    \includegraphics[width=0.48\textwidth]{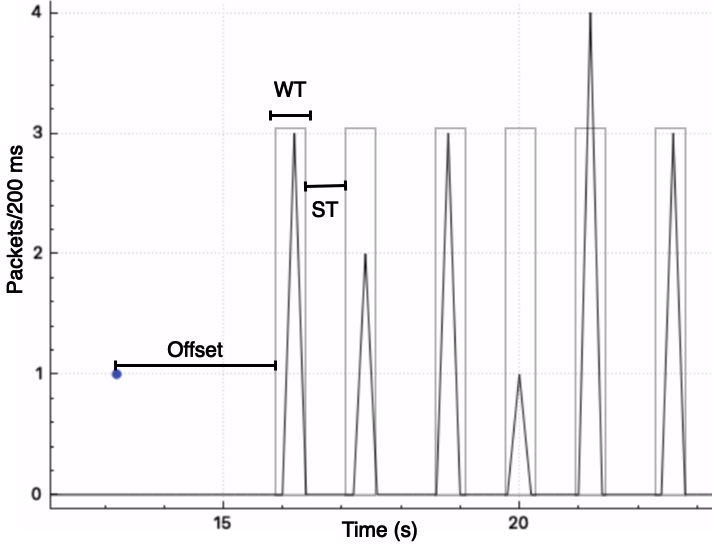}
    \caption{Wireshark I/O graph illustrating the operation of TWT.}
    \label{fig:TWT-Illustration}
\end{figure}

It is the time between the end of TWT negotiations (represented as a dot before the first spike in Figure~\ref{fig:TWT-Illustration}) and the first WT. We refer to this as \textbf{offset} for the reminder of the paper.  

A triplet of $\langle \text{WT, ST, offset} \rangle$ identifies a TWT schedule of a client.

\section{Related Work}
\label{sec: Related Work}
We present a brief survey of existing work related to scheduling in Wi-Fi and TWT in this section.

Contention-free MAC scheduling in Wi-Fi using modification to the conventional CSMA/CA mechanism has been well studied. Kuo et al.~\cite{b5} propose a synchronization mechanism using a fixed back-off value in conventional CSMA/CA to achieve a round-robin scheduling of transmission opportunities in conventional CSMA/CA. Authors in \cite{b6} propose the use of stochastic optimization problem for resource allocation taking into account variable channel conditions and arrival patterns to provide slices with diverse QoS requirements.

Schneider et al.~\cite{b4} demonstrate a proof of concept deterministic scheduling system for Industrial-IoT by using Wi-Fi 6 features of TWT and OFDMA, as an extension of the wired time-sensitive networks. A real time testbed is used to evaluate the jitter and latency, showing the potential of TWT for deterministic scheduling. In \cite{b12}, the authors propose a combination of multi-link operation and TWT based optimization for energy conservation. The work discusses the trade-off of using Broadcast TWT, which saves energy but is inefficient in high traffic scenarios, and multi-link operation, which is favorable in high traffic situations but is energy hungry. The authors propose a joint optimization approach to save energy while meeting the traffic demands in a congested scenario. Hegde et al.~\cite{b9} look into a practical implementation of downlink flow control where an intermediate proxy controls the TCP traffic by allocating time-slices to each user and setting up priorities by marking the WLAN packets differently. The work also controls the uplink TCP rate by controlling the flow of the ACKs. Sheth et al.~\cite{b16} use individual TWT to provide non-overlapping schedules to schedule traffic for bursty, constant rate, and sparse periodic IoT sensor data, and compare metrics like channel occupancy and retransmission rate along with traditional metrics. The work in \cite{b11} looks into the latency requirements of commercial internet applications. The authors classify internet traffic into categories based on latency tolerance and use state machines to classify the applications based on packet arrival patterns in real time. The authors use an approach of matching the applications to the predefined TWT intervals and also update the cycles when deemed necessary. The work provides a practical testbed example of using deterministic scheduling to protect the applications from the effects of contention. Authors in \cite{b2} test the performance of TWT for synthetic video streaming traffic in a congested scenario. They only consider a specific traffic class with the traffic congestion created using only downlink traffic, and test the performance of a single TWT client. Puligheddu et al.~\cite{b25} propose a TWT based scheduling algorithm for latency sensitive Industrial-IoT applications, using a scheduling strategy that respects the Age of Information (AoI) of the packets while reducing the net energy consumed, while demonstrating the effectiveness of TWT as a channel access mechanism in an industrial setting. However, the authors do not address the fundamental issue of congestion that largely plagues dense urban and enterprise Wi-Fi deployments and their interaction with the legacy network protocol stack and implementation issues in real life commodity hardware that severely affect the throughput of Wi-Fi. Authors in~\cite{b26} highlight the need for using time sensitive networking and its superior performance when compared to legacy CSMA/CA based channel access, while also highlighting the need to simultaneously satisfy the need of high throughput and low latency often associated with demands of interactive AR/VR applications. The authors reckon that scheduled access as a combination of OFDMA, TWT combined with smarter MAC layer scheduling is the future of Wi-Fi based networks, even in traffic agnostic scenarios.

Most of the existing work proposes changes to the scheduling algorithms or channel access mechanisms which are not practical in commercial deployments as those deviate too much from the relevant standards. Contrary to this, in our work, we propose to use a standard supported method to facilitate deterministic channel access that can be practically deployed to reduce contention and improve the net throughput performance of the system, as we demonstrate in the rest of this paper.


\section{Experimental setup}
\label{sec: Experimental setup}
\begin{figure}[!t]
\centering
    \includegraphics[width=0.48\textwidth]{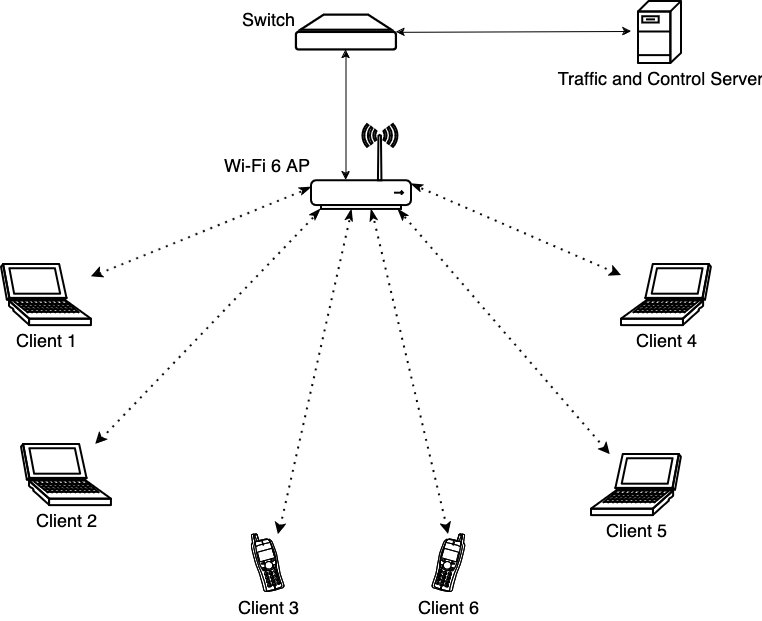}
    \caption{Schematic diagram of the testbed.}
    \label{fig:Setup}
\end{figure}

In this section, we give a brief description of the testbed used to run our experiments. A schematic representation of the test setup is shown in Figure~\ref{fig:Setup} and Table~\ref{ Test Bed Parameters} 
gives a brief description of the relevant experimental configuration.  

The maximum achievable throughput of the wireless links is found using iPerf3 \cite{b14}. The experiments are run with the parameters of individual,  trigger enabled transmissions with announced TWT commands and TWT protection disabled. All iPerf3 traffic is set as the best effort. The AP is set on full transmission power to reduce packet error rate. We choose a 20MHz clean corner channel (Channel 165), verifying that there are no other SSIDs operating on that channel over the duration of the experiments. The AP uses OpenWrt operating system \cite{b17}, which allows us to control experimental parameters like MCS index, packet aggregation and solicit TWT commands on demand. Client 1 utilizes the Qualcomm QCA639x Network Adapter. Client 2 and Client 5, use the Qualcomm FastConnect 6900 network adapter. These cards are some commercially available ones that better support TWT out of the box. Client 2 is chosen for the experimental throughput performance analysis of TWT. The network setup is not connected to the internet, to eliminate the overhead of network advertisement and management packets.

\begin{table}[!t]
\centering
    \begin{tabular}{ |p{1.2cm}||p{4.8cm}| }
        \hline
        Component & Type    \\ \hline
        Client 1 & Wi-Fi 6, \textbf{TWT-capable} laptop running Windows 10  \\  \hline
        Client 2 & Wi-Fi 6, \textbf{TWT-capable} laptop running Windows 11 \\   \hline
        Client 3 & Wi-Fi 6, Non-TWT mobile phone running Android 14 \\  \hline
        Client 4 & Wi-Fi 6, Non-TWT-capable laptop running Ubuntu 23.04\\  \hline
        Client 5 & Wi-Fi 6, \textbf{TWT-capable} laptop running Windows 11 \\   \hline
        Client 6 & Wi-Fi 6, \textbf{TWT-capable} mobile phone running Android 14 \\  \hline
        AP &  Arista C230, Enterprise grade Wi-Fi 6 AP, running OpenWrt, version 15.05.1\\    \hline
        Switch & Arista 710-P12, Used to provide L2 switching and power supply\\   \hline
        Traffic server & Linux Ubuntu 22.04 server pumping iPerf3 traffic on LAN\\  \hline
    \end{tabular}
    \caption{Description of various testbed components.}
    \label{ Test Bed Parameters}
\end{table}

\section {Performance Analysis of TWT}
\label{sec: Performance Analysis of TWT}
The  IEEE 802.11ax standard \cite{b20} uses the mantissa-exponent method to represent the values of the TWT waketime (WT) parameters mentioned in Section~\ref{sec: TWT Section}. The standard specifies the WT to be expressed as $mantissa \times 2^{exponent}$ and it can take values that are multiples of 256$\mu$s. 

As stated earlier, a TWT schedule can be represented by WT, ST, and offset combination. 
However, this representation may not be unique as a given WT can be represented in multiple ways, with different combinations of mantissa and exponent. This necessitates an unambiguous representation of TWT schedules. Hence, we define Active Airtime (AA) and Multiplication Factor (MF). The tuple $\left<AA, MF\right>$ can be used to uniquely identify a TWT schedule, as we describe below.

\subsection{Active Airtime (AA)}
\label{subsec: Active Airtime}
The IEEE 802.11ax standard mandates the maximum WT ($WT_{max}$) of a TWT client at 65280$\mu$s. This limits the maximum time a client can stay awake in a TWT schedule\footnote{This is done as per the maximum TxOP limit of legacy Wi-Fi in order to maintain interoperability.}. The Active Airtime (AA) is defined as  the percentage of time a client stays awake during a TWT schedule. 
\begin{equation*}
    \text{AA (in \%)} = \frac{WT}{WT + ST} \times 100.
\end{equation*}

Though AA(\%) may appear similar to the more familiar notion of duty cycle, it is important to note that changes is duty cycle are accommodated by changing waketimes while keeping the cycle time constant. However, this is not a viable approach in our context due to TWT's strict restrictions on $WT_{max}$ and feasible WT values.
 
\subsection{Multiplication Factor (MF)}
Though AA gives the percentage of time a client is awake in a TWT schedule, the client may achieve the same AA by waking up more (or less) frequently. Thus, we define another parameter, called Multiplication Factor (MF) to determine how frequently a client wakes up for a fixed AA during a TWT schedule, as follows:
\begin{equation*}
WT_{MF}\!\!\!\!\mod 256 = 0,
\end{equation*}
where $WT_{MF} = \left\lceil \frac{WT_{max}}{MF} \right\rceil$, $MF \in \mathbb{R}^+$. However, as the waketime (WT) can only take the values which are multiples of 256$\mu$s, the values of MF are restricted to the real interval $[1, 255]$.

\begin{figure}[!t]
\centering
\includegraphics[width=0.48\textwidth]{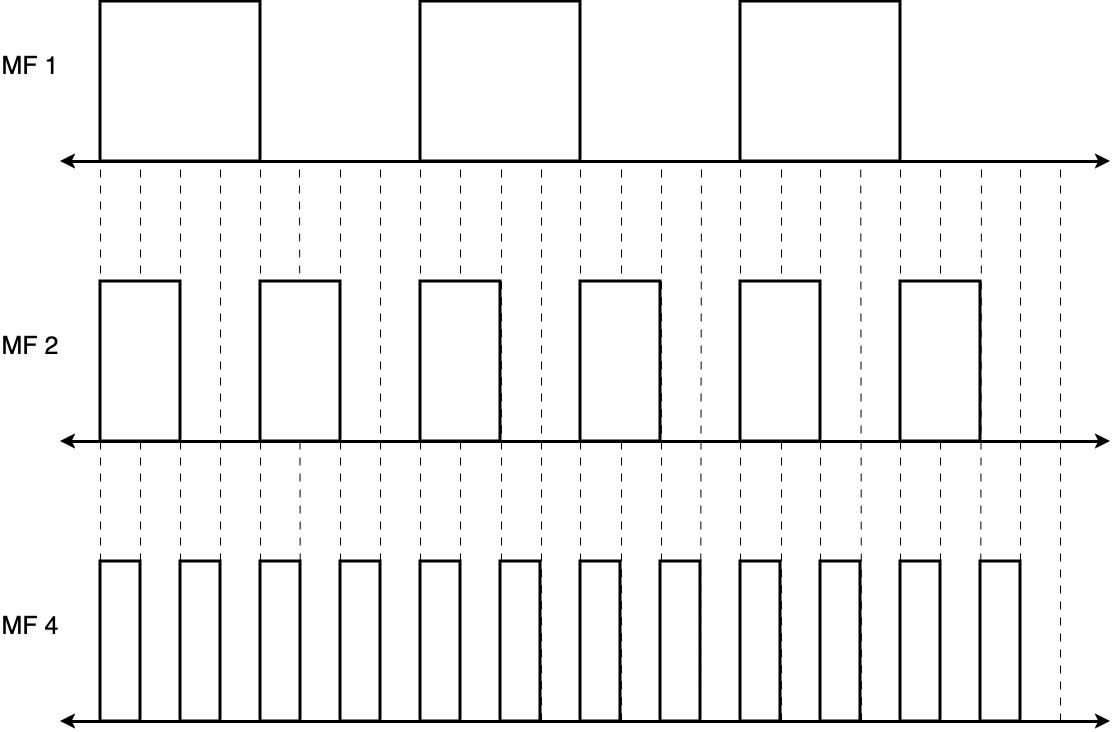}
\caption{An illustration of Multiplication Factor (MF.)}
\label{fig:TWT MF's}
\end{figure}

While MF may appear similar to frequency, it is defined with additional constraints to adhere to the rules mandated by the standard. Hence, a cycle with MF-2 wakes up and sleeps twice as often as a cycle with MF-1 for the same AA. Figure~\ref{fig:TWT MF's} shows an illustration of MFs 1, 2 and 4 for a AA of 50\%.

We quantify the effects of TWT on the application layer throughput through a series of experiments performed on our testbed described in the previous section. Traffic is pumped from the server only to Client 2 for different TWT schedules, while the rest of the setup is kept unconnected.

Throughput measured at the application layer is affected by many variables and in our experiments,  we try to quantify the effects of MAC layer packet aggregation, round trip delay (for TCP traffic) and MCS index \cite{b15} of the transmission. The MCS index is fixed for the experiments at discrete intervals (MCS 0 - MCS 11) by disabling rate adaptation schemes, this  gives us a more robust and controllable parameter, unlike RSSI, which can be easily affected by minute changes in the  experimental environment. It is also important to note that the uplink clients do not strictly adhere to TWT schedules and may choose to transmit when deemed necessary. We assume that uplink throughput trends are similar to downlink trends, though this requires further validation.

\begin{figure*}
\centering
\begin{subfigure}{1\textwidth}
\centering
    \includegraphics[width=0.8\linewidth]{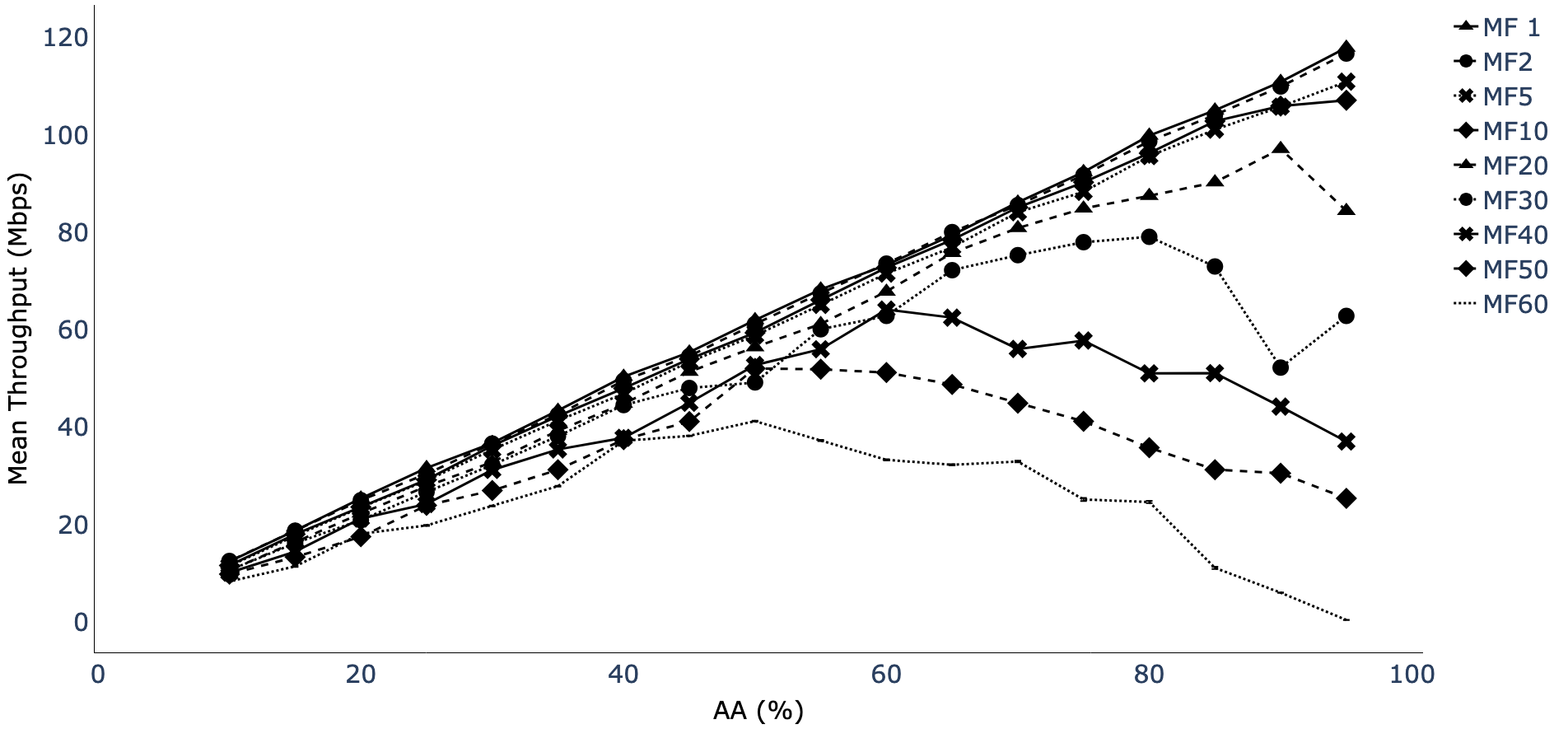}
    \caption{UDP throughput vs AA (\%)}
    \label{fig:UDP vs AA}
\end{subfigure} %
\begin{subfigure}{1\textwidth}
\centering
    \includegraphics[width=0.8\linewidth]{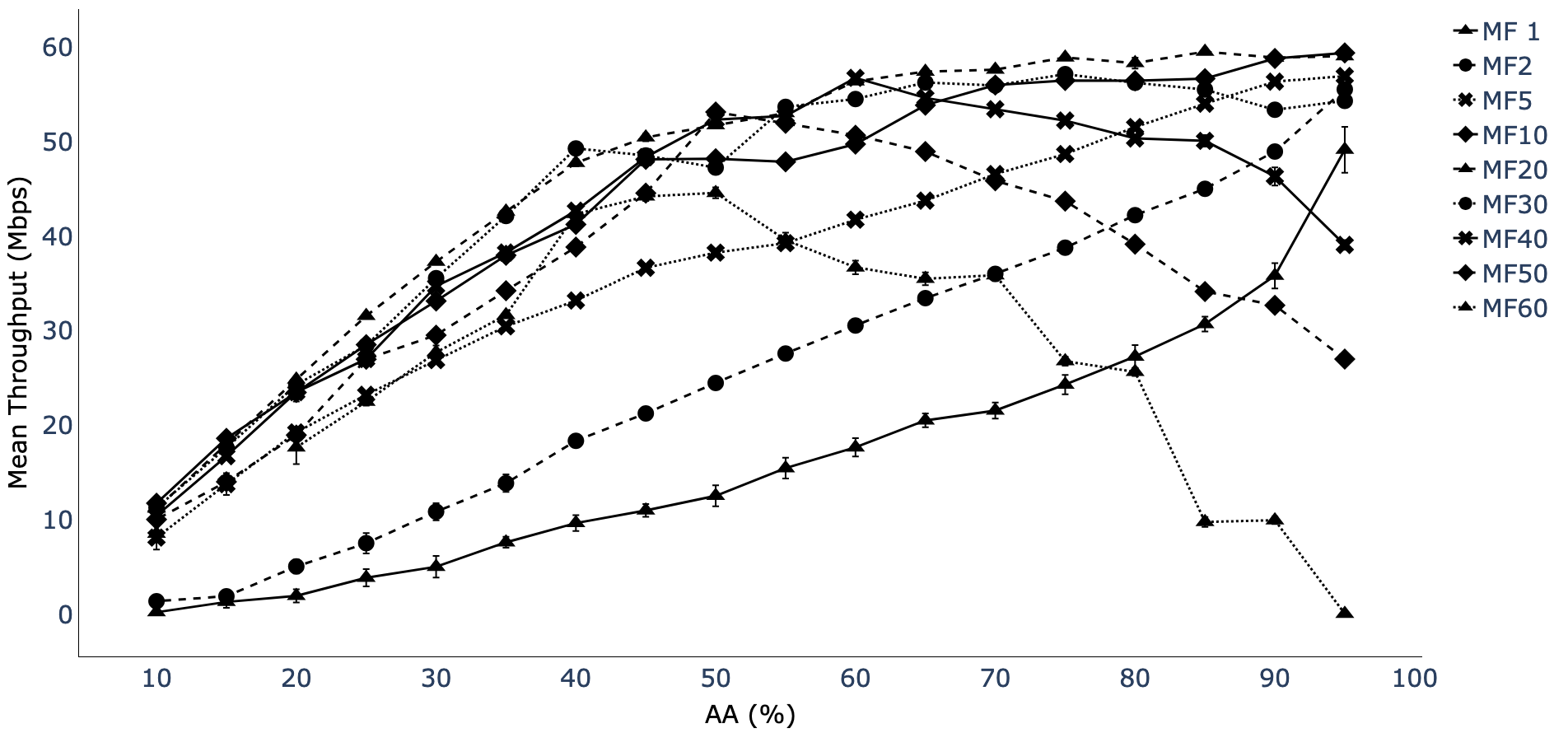}
    \caption{TCP throughput vs AA (\%)}
    \label{fig: TCP vs AA}
\end{subfigure}
\caption{ Throughput variation of TWT schedules with MCS index 7.}
\label{Througput vs AA}
\end{figure*}
     
\subsection{Throughput trends}
\label{subsec: Throughput trends}
This subsection describes the effect of TWT schedules on application throughput. A TWT-capable client is solicited with TWT schedules and its iPerf (for both UDP and TCP) throughput is measured at the server. The TWT schedules with varying MFs and AAs are solicited and their corresponding throughputs are measured.

To test UDP performance, we pump 30 seconds of iPerf UDP traffic from our traffic server (enough to saturate the buffers and cause packet drops.) The graph in Figure~\ref{fig:UDP vs AA} shows the performance of UDP traffic for different TWT schedules. Similarly, to test TCP performance, we induce a constant 40ms RTT delay for TCP BBR v1 on the server's Ethernet link using Linux `tc' commands to mimic real-world TCP connections. We choose TCP BBR as it is one of the most widely used TCP versions and is not overly reactive to packet losses, thus reducing the measurement variations. Rate adaptation is disabled to further reduce the variability in the observations. The experiment is run 30 times while keeping the MCS indices fixed. Figure~\ref{fig: TCP vs AA} shows the performance of TCP traffic for different TWT schedules.

\subsection{Factors that affect performance}
\label{subsec: Factors that affect performance}
We observe a similar trend for all MCS indices where the UDP throughput linearly increases for low MFs for all AAs, but there is a drop-off in throughput for higher MFs at larger AAs. Similarly, TCP's performance initially improves as we move from lower to higher MFs, then plateaus, and finally drops-off for larger MFs at higher AAs.

\begin{figure}[!t]
\centering
\begin{minipage}{.5\textwidth}
  \centering
  \includegraphics[width=\textwidth]{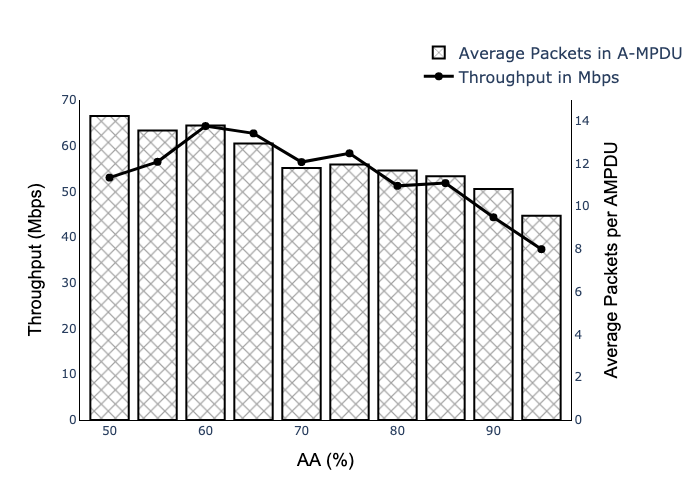}
  \captionof{figure}{Variation of packet aggregation for different AA and MF 40.}
  \label{fig:Packet Aggregation}
\end{minipage}%
\begin{minipage}{.5\textwidth}
  \centering
  \includegraphics[width=\textwidth]{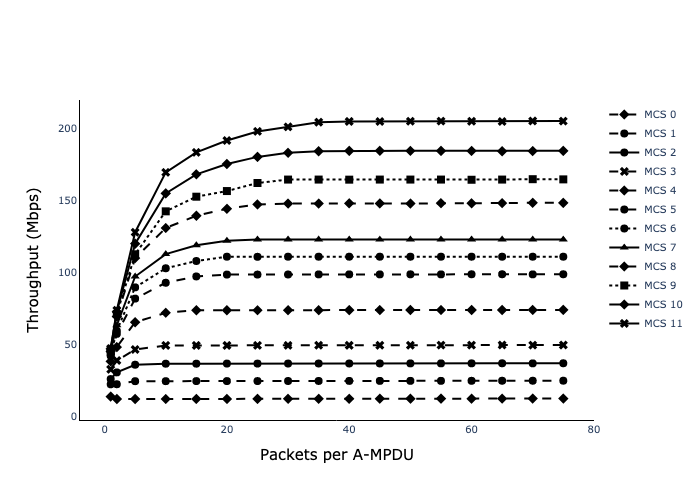}
  \captionof{figure}{Effect of packet aggregation on throughput.}
  \label{fig:Throughput vs Aggregation}
\end{minipage}
\end{figure}

The MPDU aggregation introduced in 802.11n, allows multiple packets to be encapsulated into a jumbo frame to be acknowledged with a single block ACK packet. Deeper inspection of the Wi-Fi packet-captures at higher AAs and MFs reveals that the MAC layer packet aggregation drops off at higher MFs. This increases the overhead associated with packet transfer including Short Inter Frame Space (SIFS) and ACK transmit times (which are transmitted at legacy rates.) Figure~\ref{fig:Packet Aggregation} shows the average packet aggregation and throughput achieved for different AAs in the range of 50-95\%, It clearly shows a strong relationship between packet aggregation and throughput. Figure~\ref{fig:Throughput vs Aggregation} shows the results of changing the number of packets aggregated into a single MPDU, while keeping the MCS index fixed. This shows that packet aggregation is not the primary factor that determines the throughput behavior, as the aggregated MPDUs should lead to a higher throughput than that observed while using TWT.

\begin{figure}[!t]
\centering
\begin{minipage}{.5\textwidth}
  \centering
  \includegraphics[width=\textwidth]{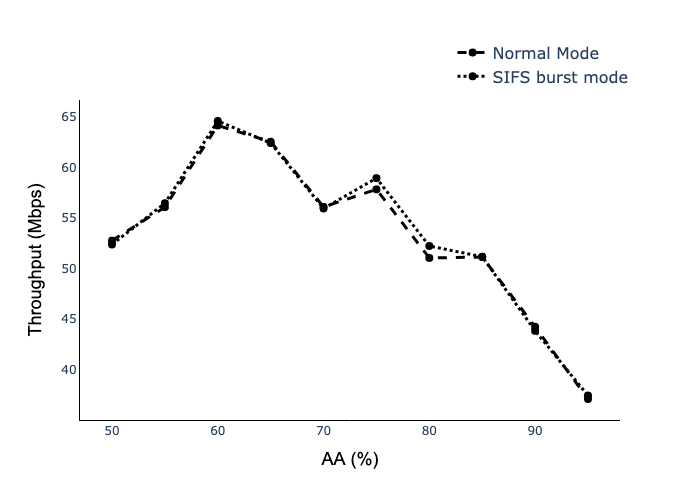}
  \captionof{figure}{Effect of SIFS burst mode on throughput.}
  \label{fig:SIFS burst mode}
\end{minipage}%
\begin{minipage}{.5\textwidth}
  \centering
  \includegraphics[width=\textwidth]{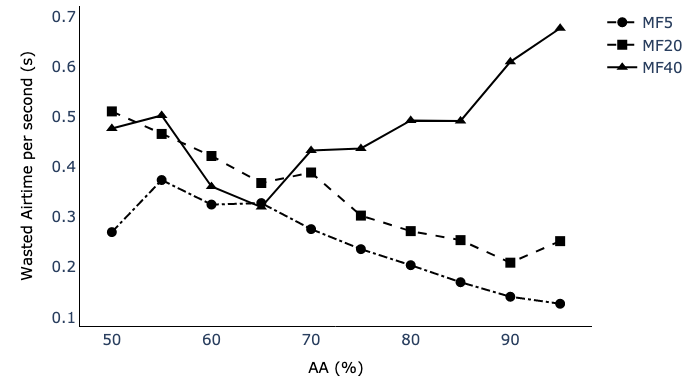}
  \captionof{figure}{Variation in the wasted airtime for different MFs.}
  \label{fig:Wasted Airtime}
\end{minipage}
\end{figure}

This observation is further substantiated by setting the AP in SIFS burst mode, which allows the transmission of multiple frames within a single TxOP. In this mode, multiple frames are burst with only a SIFS between them~\cite{b23}. Enabling SIFS burst mode showed no discernable difference in the throughput performance as shown in Figure~\ref{fig:SIFS burst mode}.

We also observe increased wasted airtime or the average idle airtime per second during which there is no packet transmissions at higher AAs and MFs (Figure~\ref{fig:Wasted Airtime}). This trend of increase in wasted airtime for higher AAs is observed only at higher MFs, whereas lower MFs show a decrease in the wasted airtime with corresponding increase in throughput for higher AAs. It is also worth noting that during a single WT duration, only one MPDU frame is transmitted and the remaining airtime is idle. 

We hypothesize that at higher AAs and MFs the WT and ST are smaller. Therefore, with a higher operating frequency of the TWT schedules, the penalty due to wasted airtime for every wake-sleep cycle adds up, affecting the overall throughput of the TWT schedule. This may require further analysis and we plan to take it up in our future work.

\section{The Optimization Problem}
\label{sec: Optimization Problem}
From the performance analysis of TWT in last section, it is clear that the throughput of the clients is primarily dependent upon the TWT schedules and intuitively, overlapping schedules may affect the overall system throughput. In this section, we formulate an optimization problem to provide optimal WT and ST allocations to all the clients connected with an AP in a multi-client Wi-Fi 6 system, such that the total throughput of the Basic Service Set (BSS), the wireless network formed by the AP and associated clients, is maximized. The resultant time allocations are then converted to TWT schedules according to Section~\ref{sec: TWT Section} by setting appropriate TWT offsets.

We also take into the account the issues that occur with the CSMA/CA channel access, including the uplink/downlink disparity, and provide minimum throughput guarantees for the clients that need to be protected. It is important to note that we only provide statistical throughput guarantees, not guarantees for individual data packets or flows, though transmission and proportional throughput fairness are emphasized.

We formalize these objective and constraints in the following optimization problem using the notation in Table~\ref{table:notation}. Then, in the rest of the paper, we discuss challenges in solving it, the proposed solution approach, and its numerical and experimental evaluation.

\begin{table}[!t]
\centering
\begin{tabular}{ |p{1.2cm}||p{4.8cm}| }
    \hline
    Notation    &   Meaning  \\    \hline
    $\mathcal{N}$   &     Set of all clients \\ \hline
    $T$   & Throughput using TWT  \\    \hline
    $T_{Th}$   & Guaranteed throughput  \\    \hline
    $T_{CSMA}$  & Throughput using CSMA/CA    \\   \hline
    $\mathcal{K}$   & Set of clients with protected throughput  \\    \hline
    $\mathcal{D}$   & Set of unprotected downlink clients   \\   \hline
    $\mathcal{U}$   & Set of unprotected uplink clients \\     \hline
    $WT_{i}$  & Waketime of the $i^\textrm{th}$ client \\  \hline
    $ST_{i}$  & Sleeptime of the $i^\textrm{th}$ client \\   \hline
    $C$   & Maximum MF of any schedule \\    \hline
    $T_{avg}$ & Average throughput of downlink vanilla clients\\    \hline
    $\mathcal{A}$ & Set of sampled AAs  \\  \hline
    $\mathcal{M}$ & Set of sampled MFs  \\    \hline
\end{tabular}
\caption{Notation}
\label{table:notation}
\end{table}

\begin{center}
    \begin{subequations}
    \begin{equation}
        \label{eqn: Proportionally fair}
        Maximize\;\sum_{i=1}^{n}\log(1+ T_i),
    \end{equation}
    \begin{equation}
        \label{eqn: Minumum throughput gaurantees}
     \mbox{such that }\,\,\,  T_i \geq T_{th,i}, i \in \mathcal{K},
    \end{equation}
    \begin{equation}
        \label{eqn: Better than CSMA/CA}
        T_i \geq T_{CSMA,i}, i \in \mathcal{D},
    \end{equation}
    \begin{equation}
        \label{eqn:Uplink downlink fairness}
        T_i \leq T_{avg}, i \in \mathcal{U},
    \end{equation}
    \begin{equation}
        \label{eqn: enforce round-robin}
        ST_i = \sum _{j \in \mathcal{N}, j \neq i} WT_j, i \in \mathcal{N},
    \end{equation}
    \begin{equation}
        \label{eqn: MF cost}
        \frac{WT_{max}}{WT_i} \leq C, i \in \mathcal{N},
    \end{equation}
    \begin{equation}
        \label{eqn: Function definition}
        T_i= f(WT_i,ST_i,\textit{MCS Index(i)}), i \in \mathcal{N},
    \end{equation}
    \begin{equation}
        \label{eqn: Service period domain}
        WT_{i} \leq WT_{max}\text{ and } WT_{i} = 256 \times k, k \in \mathbb{N}, i \in \mathcal{N},
    \end{equation}
    \begin{equation}
        \label{eqn: Wake interval domain}
        WT_{i} \geq 0, i \in \mathcal{N}.
    \end{equation}
\end{subequations}
\end{center}

The objective in Eqn.~\ref{eqn: Proportionally fair} maximizes the throughput of the system. We use the logarithmic utility function that ensures proportionally fair throughput allocation to the clients (both uplink and downlink) with varying MCS indices. This ensures that the higher MCS clients do not receive a disproportionate fraction of the available airtime, thus imposing a notion of macro-level fairness. To maintain application level QoS, we provide a minimum throughput guarantee to clients. The set of TWT clients and their associated throughput requirements are represented by the sets $\mathcal{K}$ and $T_{th}$, respectively, and the corresponding constraints represented by Constraint~\ref{eqn: Minumum throughput gaurantees}. Constraint~\ref{eqn: Better than CSMA/CA} ensures that the throughput performance of the clients are better than the baseline CSMA/CA throughput for downlink scenario, while Constraint~\ref{eqn:Uplink downlink fairness} limits the throughput of uplink clients to the average of downlink clients reducing the uplink/downlink disparity.

\begin{figure}[!t]
\centering
    \includegraphics[width=0.45\textwidth]{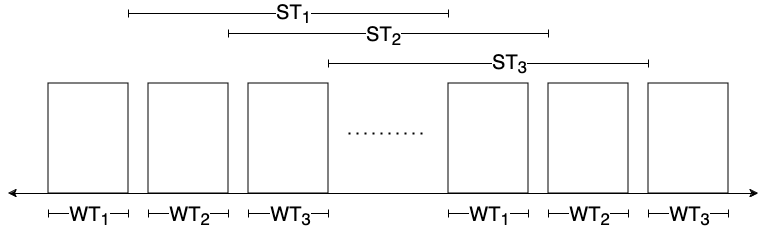}
    \caption{An example of scheduling under round-robin conditions and the relationship between the WTs and STs.}
    \label{fig:Round Robin}
\end{figure}

Further, to address the issue of micro-fairness, we ensure round-robin scheduling  as described in \cite{b8}. We enforce the ST of each client to be at least equal to the sum WT of all other clients as described in Constraint~\ref{eqn: enforce round-robin} (Figure~\ref{fig:Round Robin}), ensuring that every client has an opportunity to transmit before the next cycle begins. Constraint~\ref{eqn: MF cost} restricts the MFs to a maximum limit as there is a cost associated with waking up and sleeping more frequently. Constraint~\ref{eqn: Function definition} uses a hitherto unspecified analytical function that relates the throughput achieved by a TWT schedule to various system parameters. In the next section, we provide a method to define such a function. Finally, Constraints~\ref{eqn: Service period domain} and \ref{eqn: Wake interval domain} mandate the schedules to be compliant with the requirements described in Section~\ref{sec: Performance Analysis of TWT}, which limit the WT of the schedule to $WT_{max}$ and ensure that it is a multiple of 256$\mu$s.

\section{The Proposed Solution approach}
\label{sec: Solution approach}
In order to solve the optimization problem \eqref{eqn: Proportionally fair}, we need a relationship between the TWT schedules of all clients and the corresponding throughput (Constraint~\ref{eqn: Function definition}). In Section~\ref{sec: Performance Analysis of TWT}, we argued that a TWT schedule for every client can be uniquely characterized using the corresponding $\left<AA, MF\right>$ tuple. However, as both AA and MF are real-valued, a client's optimum schedule is found by searching over the space $\mathbb{R}^2$ for the corresponding optimum $\left<AA, MF\right>$ tuple and this results in the search for an optimum schedule for $n$ clients over $\mathbb{R}^{2n}$ - an infinitely large solution space. However, given the highly non-convex nature of the solution space, partially resulting from the constraints of non-overlapping and round-robin schedules, no efficient solution scheme is available that may guarantee the globally optimal solution and insensitivity to initial value.

To reduce the size of the solution space, we first restrict both AA and MF to only discrete values. The set $\mathcal{A}$ of AAs is restricted to 18 values in $[10, 95]$ at a granularity of 5\% and the set $\mathcal{M}$ of MFs is restricted to only 255 values, such that $\frac{WT_{max}}{MF} \mod 256 = 0$. Further, we experimentally observe that MF values greater than 60 lead to deterioration in the system throughput, so we further restrict the set $\mathcal{M}$ to 21 such values that satisfy both conditions on MF values. Using these 18 values of AA and 21 values of MF, a client can construct 378 unique corresponding TWT schedules. In fact, for each of these schedules, we can precompute the corresponding throughput and store it in a table along with the corresponding $\left<AA, MF\right>$ tuple. This allows us to replace the function computation in Constraint~\ref{eqn: Function definition} by a table look-up. For $n$ clients, this still leads to an exponentially large solution space, that is, of size $378^n$. In subsection~\ref{subsec: Reducing the runtime} below, we propose a method to reduce the runtime to \textit{reasonable} values for small WLANs, consisting of up to 10 clients.

Using the above arguments, the function in Constraint~\ref{eqn: Function definition} is modified to convert our optimization problem to a combinatorial optimization problem, by using the following equations, where every client $i$ is associated with a throughput vector $T_{sample,i}$, containing the throughput achieved for the MF $m$ and the AA $a$. So, $\forall a \in \mathcal{A}$, $\forall m \in \mathcal{M}$, $i \in \{0,1\}$ and $n = |\mathcal{A}| \times |\mathcal{M}|$, define
\begin{align*}
    \centering
    T_{sample,i}&=\left[T^{MCSi}_{a,m}\right]_{n \times 1},\\
    WT_{sample,i}&=\left[WT_{a,m}\right]_{n \times 1},\\
    ST_{sample,i}&=\left[ST_{a,m}\right]_{n \times 1},\\
    I_{i}&=\left[i_{a,m}\right]_{{n \times 1}}, [1]^n \times I_i^{\intercal}=1,
\end{align*}
where the vectors $WT_{sample} \text{ and } ST_{sample}$ contain the corresponding WTs and STs, and $I_i$ is the indicator variable vector used to pick the values according to the following set of equations: 
\begin{align*}
    T_i=T_{sample,i} \times I_i^{\intercal}, \\ 
    WT_i=WT_{sample,i} \times I_i^{\intercal}, \nonumber\\
    ST_i=ST_{sample,i} \times I_i^{\intercal}. \nonumber
\end{align*}

Hence, the relationship between the schedules and the corresponding throughputs, outlined by Constraint~\eqref{eqn: Function definition}, is modified to picking the appropriate indicator variable vector for each client, while satisfying the throughput requirements outlined by Constraints~\ref{eqn: Minumum throughput gaurantees},~\ref{eqn: Better than CSMA/CA} and~\ref{eqn:Uplink downlink fairness}. Selection of the indicator variable should respect the round-robin criterion in Constraint~\ref{eqn: enforce round-robin}. Since WT and ST correspond to the actual TWT commands used in experimentation, they automatically satisfy the Constraints~\ref{eqn: MF cost},~\ref{eqn: Service period domain}, and~\ref{eqn: Wake interval domain}.
  
\subsection{Issues with this approach}
\label{subsec: Issues with this solution}
In the numerical solutions of the optimization problem with the solution approach proposed above, we observe two issues.
First, with the combination of limited sample points, logarithmic utility function, and strict conditions of round-robin scheduling, the solution always tends to be the same TWT schedule repeating for all the clients, i.e. equal division of airtime for all clients (regardless of MCS indices, with the  minimum throughput criterion satisfied). Second, the non-overlapping TWT schedule condition along with the round-robin scheduling forces a single client to dictate the TWT schedule for all others (generally, the client with the highest MCS index). This introduces an undesirable dominance aspect in the system as the client with the highest MCS index may dictate the MF and AA of all other clients in the system.

Both issues can be addressed by one of the two approaches: including appropriate constraints in the above optimization problem or letting the optimization problem be as it is and applying the constraints corresponding to these two issues on its resulting solution. Technically, there is no difference between these two approaches, but the first approach, though it appears neat, is somewhat opaque in that it does not provide much insights into the application of the corresponding constraints and resulting solution, while the second approach lets us do so. Therefore, we adopt the second approach in the rest of the paper and introduce the concepts of ``pseudo client'' and ``acceptable overlap'' next to implement it.

\subsection{Pseudo client and acceptable overlap}
\label{subsec: Pseudo Client approach and Acceptable Overlap}
\begin{figure*}
\centering
    \includegraphics[width=0.8\textwidth]{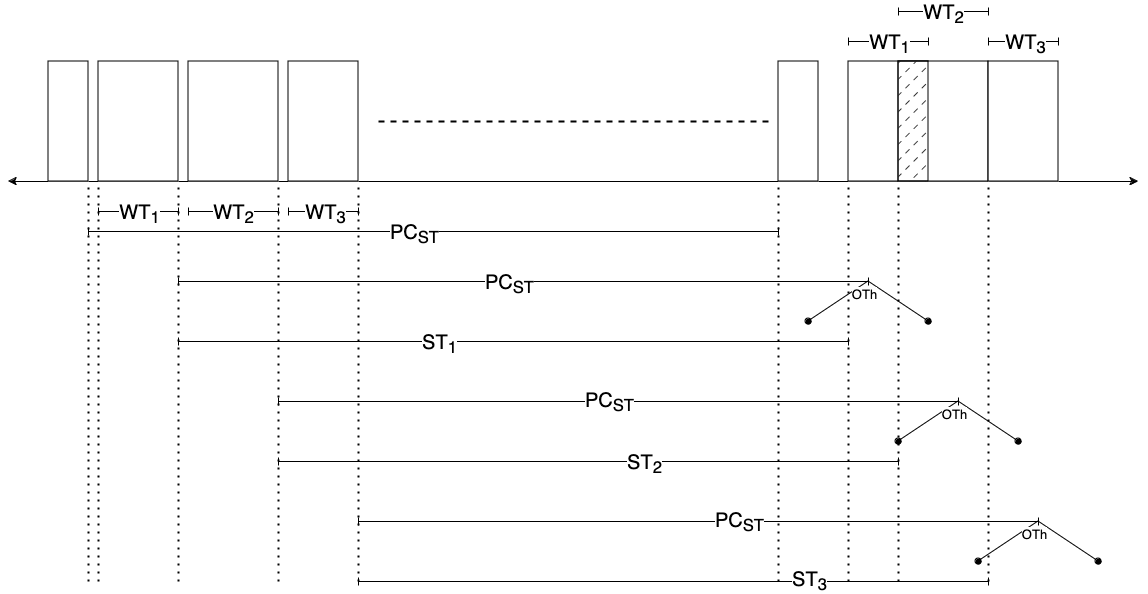}
    \caption{An illustration of how $PC_{ST}$ and $OTh$ affect TWT schedules. The dashed area represents the overlap between the waketimes $WT_{1}$ and $WT_{2}$.}
    \label{fig: Pseudo Client}
\end{figure*}

In order to address the above-mentioned issues, we propose a ``pseudo client'' approach, where a virtual client has the highest priority and dictates the ST of all the clients. The pseudo client's WT does not contribute to the utility in the objective function, so it can be viewed as wasted airtime, requiring its minimization. The pseudo client's sleep time ($PC_{ST}$) now dictates the WT and ST of all other clients (Figure~\ref{fig: Pseudo Client}). Since we want to avoid the situation where all the clients are forced to use the same schedule, below we introduce a parameter, ``overlap threshold'' $OTh$, defined in terms of how much deviation a client's ST can have from the ST of the pseudo client, instead of all clients being forced to have the same sleep value.

\begin{align*}
    \label{eqn: Overlap Threshold}
    |{PC}_{ST}-ST_i| \leq OTh, \forall i \in \mathcal{N}.
\end{align*}
Hence, the round-robin scheduling requirement of Constraint~\ref{eqn: enforce round-robin} is modified to,
\begin{equation*}
    {PC}_{ST} \geq \sum_{i \in \mathcal{N}} WT_i.
\end{equation*}

The threshold, $OTh$, allows some overlap in the TWT schedules of the clients. To give bound on the loss of throughput due to the overlap, we assume that every time there is an overlap among clients, the one with the lowest MCS index always wins the right to transmit. This allows us to compute a lower bound on the throughput achieved by a TWT schedule.

The Algorithm~\ref{alg: Calculate Overlap Loss } describes how the loss due to overlap is computed. The $ListofIntervals$ structure stores the TWT wake-up instances as intervals with start time, end time, identity of the client $i$ associated with the interval, and finally, the MCS of the transmission. The function $indexof()$ returns the index of the client with the maximum MCS index, among the clients being compared. In Algorithm~\ref{alg: Calculate Overlap Loss }, the throughput loss due to overlap is calculated for each client over one second period and the loss is then subtracted from the corresponding throughput achieved by the TWT schedule in the absence of overlap to compute the final throughput of the client $i$. This throughput is then used to evaluate the objective function \eqref{eqn: Proportionally fair}, while ensuring the constraints in \ref{eqn: Minumum throughput gaurantees}-\ref{eqn:Uplink downlink fairness} remain satisfied.
\begin{algorithm}
    \caption{Loss due to overlaps}
    \label{alg: Calculate Overlap Loss }
    \begin{algorithmic}
    \Require  $\{WT_i :\forall i \in \mathcal{N}\}  \text{ and } \{ST_i : \forall i \in \mathcal{N}\}$
    \Ensure $WT_i \neq 0 \text{ and } ST_i \neq 0 \text{ } \forall i \in \mathcal{N}$
    \State $start_i=\sum_{j < i, j \in \mathcal{N}}WT_j$
    \State $end_i=start_i+WT_i$
    \State $ListofIntervals=[null]$ \Comment{Stores TWT interval}
    \For{$ i \in N$}
        \While{$end_i \leq 1000000$}
            \State $ListofIntervals.append(Start_i, end_i,i,MCS_i)$
            \State $start_i= start_i+ST_i$
            \State $end_i=start_i+WT_i$
        \EndWhile
    \EndFor
    \State $sort(ListofIntervals)$ \Comment{Sort according to end}
    \For {$\forall i \in ListofIntervals$}
        \For {$\forall j \in listofIntervals \text{ and } j=i$}
            \If {$start_j < end_i$}
                \State $ Overlap=start_j - end_i $ 
                \State $index=indexof(max(MCSi,MCSj))$
                \State $OLoss_{index}= overlap \times T^{persecond}_{index}$
                \State $T_{index}= T_{index} - OLoss_{index}$
            \EndIf
        \EndFor
    \EndFor
    \end{algorithmic}
\end{algorithm}

\begin{figure}[!t]
\centering
\includegraphics[width=0.5\textwidth]{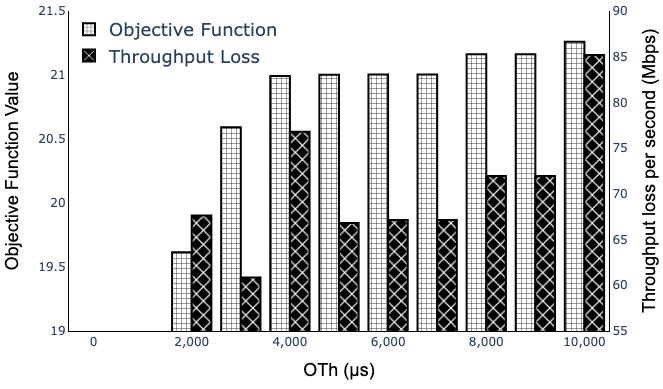}
\caption{Variation in the objective function value for different $OTh$ and its impact on the system throughput.}
\label{fig:Results}
\end{figure}

The parameter, $OTh$, can be used as a controller - by varying it the overlap among the clients schedules can be adjusted. The graph in Figure~\ref{fig:Results} shows the change in the objective function value for different $OTh$ values, and the associated loss of throughput, incurred as a result of overlap among the schedules. No feasible TWT schedules are found for $OTh$ values less than 2000$\mu$s. Increasing the $OTh$ does not always result in a better schedule. For example, the TWT schedule remains the same for $OTh$ of $6000\mu$s and $7000\mu$s, and $8000\mu$s and $9000\mu$s.

\subsection{Reducing the runtime}
\label{subsec: Reducing the runtime}
Since the loss due to overlap depends not only on a client's TWT schedule, but also on the schedules of all other 
clients and their MCS indices, the search space for the optimum TWT schedule, as argued earlier, is exponentially large. Hence, in the following, we propose an approach to reduce the runtime for small networks.

Since we need to iterate over the search space and search for each client is independent of the other, we can reduce the runtime by searching the space concurrently. The choice of the $PC_{ST}$ value dictates the sleep values of all other clients, hence reducing the search space of the clients by restricting the respective feasible sleep values to the range $[PC_{ST}-OTh, PC_{ST}+OTh]$ for each parallel search.

Further, we reduce the runtime by increasing the granularity of the MFs (Constraint~\ref{eqn: MF cost}) and the associated AAs of the sampled set in successive iterations. By searching the feasible values of protected uplink clients in the end, based on Constraint~\ref{eqn:Uplink downlink fairness}, reduces the search space based on the values of unprotected best-effort downlink clients chosen previously.

\section{Results}
\label{sec: Results}
In this section, we compare the overall system throughput achievable over our testbed, as described in Section~\ref{sec: Experimental setup}, in two scenarios: (a) when all clients use CSMA/CA and (b) when some client are TWT-capable and follow a deterministic schedule. To do so, we deploy a three step method. We configure all six clients in the testbed as mentioned in the second and third columns of Table~\ref{Table: CSMA vs TWT}\footnote{Commercial devices do not allow setting MCS index values. So, we assume the MCS index to be the statistically significant MCS index of the transmitted packets.}. In the first step, all clients are configured as CSMA/CA clients and their respective throughputs are obtained, as tabulated in the fourth column of Table~\ref{Table: CSMA vs TWT}. In the second step, using these CSMA/CA throughput values, the optimization problem in Section~\ref{sec: Optimization Problem} is solved in such a manner that the resultant throughputs, as in the fifth column of Table~\ref{Table: CSMA vs TWT}, better the CSMA/CA throughputs and the corresponding optimum TWT schedule and the system parameters are obtained. The optimization problem is solved with granularity of 5\% for the AA in the input lookup table and the maximum MF for the schedules is restricted to 40. Finally, in the third step, four TWT-capable clients are configured using these parameters values and their channel access is scheduled as per the optimum TWT schedule obtained in the second step. Their respective throughputs are obtained along with the throughputs of the remaining two CSMA/CA clients and compared with the corresponding CSMA/CA throughputs of the first step. Table~\ref{Table:All experimental data} contains the results of these comparisons. These comparisons are carried out over different days and at different times of a day to account for the possibility of different channel conditions. Each experiment is run 50 times, with each run containing 10 seconds of iPerf traffic, pumped in parallel from/to the clients. The throughputs corresponding to each run are averaged and listed in the table.
\begin{table}[!t]
\centering
\begin{tabular}{|>{\centering\arraybackslash}m{0.75cm}|>{\centering\arraybackslash}m{2.5cm}|>{\centering\arraybackslash}m{0.5cm}|>{\centering\arraybackslash}m{1.55cm}|>{\centering\arraybackslash}m{1.35cm}|}
        \hline
        Client No.& Type &  MCS & Throughput achieved in CSMA/CA (in Mbps) & Optimization problem solution (in Mbps)\\
        \hline
        1& Uplink, iPerf UDP, Protected &7&22.47&22.70\\
        \hline
        2& Uplink, iPerf UDP, Best effort &11&32.97&38.52\\
        \hline
        3&Uplink, iPerf TCP BBR with 40ms RTT, Protected & 8&21.96&30.55\\
        \hline
        4&Downlink, iPerf UDP, Protected&7&18.5&42.32\\
        \hline
        5&Downlink, iPerf UDP, Protected &7&52.37&42.32\\
        \hline
        6& Downlink, iPerf TCP BBR with 40ms RTT, Protected & 8 & 7.46&29.81\\
        \hline
\end{tabular}
\caption{Summary of the comparison of the CSMA/CA and simulated TWT performance when the $OTh$ is set to $10000\mu$s.}
\label{Table: CSMA vs TWT}
\end{table}

\begin{table*}[ht]
\centering
    \begin{tabular}{|c|c|c|c|c|c|c|c|c|c|c|}
    \hline
    \multirow{3}{*}{Client No.} & \multicolumn{10}{c|}{Experiment No.}  \\ \cline{2-11} 
    & \multicolumn{2}{c|}{ 1} & \multicolumn{2}{c|}{ 2} & \multicolumn{2}{c|}{3} & \multicolumn{2}{c|}{4} & \multicolumn{2}{c|}{5} \\\cline{2-11}
    & CSMA/CA&TWT&CSMA/CA&TWT&CSMA/CA&TWT&CSMA/CA&TWT&CSMA/CA&TWT\\ \cline{1-11}
        1&	22.47&	19.42	&26.36&	21.24	&21.23&	17.64	&20.5&	19.27&	20.36&	15.42	\\ \hline
        2&  32.97&	25.79&	32.87&	26.51&	28.05&	29.55	&60.66&	23.58	&31.51&	18.85	\\ \hline
        3&	21.96&	15.03&	10.5&	12.42	&20.83&	16.54&	9.74	&11.54&	20.96	&11.05	\\ \hline
        4&	18.5&	35.96&	19.39&	41.72	&19.17	&39.54	&14.77	&45.47	&17.56	&36.76	\\ \hline
        5&	52.37	&41.6&	45.44&	38.18&	52.01&	37.1&	39.44	&46.32	&53.73	&42.42	\\ \hline
        6&	7.46&	18.02&	8.36&	26.5&	10.2&	20	&1.53&	15.05&	7.51&	24.32	\\ \hline
        \textbf{Total}	&\textbf{155.73}	&\textbf{155.82}	&\textbf{144.92}&	\textbf{166.57}	&\textbf{151.49}&	\textbf{160.37}&	\textbf{146.64}	&\textbf{161.23}	&\textbf{151.63}&	\textbf{148.82}	\\ \hline
\end{tabular}
\caption{Testbed comparison of CSMA/CA and TWT throughputs (in Mbps) under different channel conditions.}
\label{Table:All experimental data}
\end{table*}

It is observed that overall system throughput improves in most of the scenarios and on average, when there are some TWT-capable clients using deterministic scheduling. This result is particularly interesting because it is obtained on a testbed with consumer-grade off-the-shelf components, comprising of TWT and non-TWT-capable clients, the TWT-capable clients having minimal firmware support, with a mixture of heterogeneous traffic and signal strengths (MCS indices). It will be interesting to repeat these experiments with better firmware support for TWT. 

In our experiments, we also observe that the system throughput performance deteriorates at larger MFs and this serves as a motivation to restrict the values of MF in Section~\ref{sec: Solution approach}. The Table~\ref{Table:Performance Drop at higher MFs}, illustrates one such scenario, where we observe a drop-off in the throughputs attained by the clients with higher MF schedules. It should be noted that the non-TWT clients generally perform better in this scenario. This demonstrates a scenario, where the number of collisions with the non-TWT-capable clients increases at higher MF, resulting in deteriorated throughput for TWT-capable clients, which have limited window for competing for the channel, unlike the non-TWT capable clients, which are free to compete whenever.

Further, it is also observed that solutions of the optimization problem always tends to keep the values of the TWT schedules of different clients as similar as possible (in terms of both MFs and AAs). Hence, in our experiments, we set the schedule of all the TWT-capable clients with $\langle25\%,1\rangle$, as our testbed contains only four TWT-capable clients.
\begin{table}[t!]
    \centering
    \begin{tabular}{|c|c|c|c|}
    \hline
    Client No. & CSMA/CA Throughput & TWT Schedule & TWT Throughput \\ \hline
    1 & 36.87 & $\left\langle 25, \text{MF-10} \right\rangle$ & 14.13 \\ \hline
    2 & 42.29 & $\left\langle 25, \text{MF-10} \right\rangle$ & 28.15 \\ \hline
    3 & 5.59  & N/A & 17.65 \\ \hline
    4 & 18.2  & N/A & 29.65 \\ \hline
    5 & 44.56 & $\left\langle 25, \text{MF-10} \right\rangle$ & 19.05 \\ \hline
    6 & 21.76 & $\left\langle 25, \text{MF-10} \right\rangle$ & 18.82 \\ \hline
    \end{tabular}
    \caption{Performance comparison of CSMA/CA and TWT with higher MF schedules.}
    \label{Table:Performance Drop at higher MFs}
\end{table}

\subsection{Skewed TWT schedules}
\label{subection: Skewed TWT schedules}
Previous results are obtained while attempting to maximize the overall system throughput. However, there may be scenarios where some clients need differentiated/prioritized service. Table~\ref{Table: Variation of TWT Schedule} shows that with TWT such scenarios can be addressed using skewed schedules. It also highlights that meeting requirements of such prioritized clients may result in worse overall system performance compared to
when we divide available airtime equally among the TWT-capable clients (Table~\ref{Table: CSMA vs TWT}).

Variation 1 in the Table~\ref{Table: Variation of TWT Schedule} shows a skew with all the clients having different schedules, It also highlights some of the complexities associated with providing schedules 
as clients now require varied MF requirements to satisfy the non overlapping conditions highlighted in Section~\ref{sec: TWT Section}. Variation 2 provides the same schedule to 3 TWT capable clients while prioritizing a single client. We can clearly see improved performance for the prioritized Client 5, though the overall system throughput suffers. Variation 3 is just a different permutation of the schedules in Variation 1.
\begin{table}[!t]
    \centering
    \renewcommand{\arraystretch}{1.3} 
    \begin{tabular}{|c|c|c|c|c|c|c|c|c|}
    \hline
    \multirow{2}{*}{Variations} & \multicolumn{6}{c|}{Client No.}& \multirow{2}{*}{\textbf{Total T'put}} \\ \cline{2-7}
     & 1&2&3&4&5&6& \\
    \hline
    \multirow{2}{*}{1} & 23.84 $\left<25, MF-1\right>$ & 14.9 $\left<25, MF-1\right>$&3.46$\left<25, MF-1\right>$& 18.61&37.5 $\left<25, MF-1\right>$&30.91&129.22\\\cline{2-8}
    & 19.51 $\left<30,MF-1.25\right>$& 27.39 $\left<35,MF-1\right>$ &5.48 $\left<17,MF-2\right>$&17.08&26.56 $\left<18,MF-1.9\right>$& 32.42 &128.44 \\\hline
     \multirow{2}{*}{2} & 17.06 $\left<25, MF-1\right>$ & 21.74 $\left<25, MF-1\right>$ & 5.86 $\left<25, MF-1\right>$& 28.79 & 27.2 $\left<25, MF-1\right>$& 38.59 & 139.24 \\\cline{2-8}
    & 20 $\left<20, MF-2\right>$ & 17.48 $\left<20, MF-2\right>$& 6.67 $\left<20, MF-2\right>$ & 27.76 & 37.49 $\left<40, MF-1\right>$& 20.98 & 130.38 \\ \hline
     \multirow{2}{*}{3} &24.19 $\left<25, MF-1\right>$ &18.93 $\left<25, MF-1\right>$ &3.3 $\left<25, MF-1\right>$&19.46 &26.72 $\left<25, MF-1\right>$&51.72 &144.32\\ \cline{2-8}
    & 22.96$\left<17,MF-2\right>$& 19.73 $\left<35,MF-1\right>$& 4.21 $\left<30,MF-1.25\right>$ & 13.2 & 16.94 $\left<18,MF-1.9\right>$ & 64.67 & 141.71\\ \hline
    \end{tabular}
    \caption{Throughput performance comparison of prioritized versus non-prioritized client scheduling for three different variations of client prioritization. For each variation, the top row indicates the per client throughput when all TWT-capable clients are treated equally, and the bottom row indicates the per client throughput when one or more clients are prioritized.}
    \label{Table: Variation of TWT Schedule}
    \end{table}

\subsection{Buffer Bloat and Latency}
\label{subsection: Buffer bloat}

\begin{figure}[!t]
\centering
\includegraphics[width=0.7\textwidth]{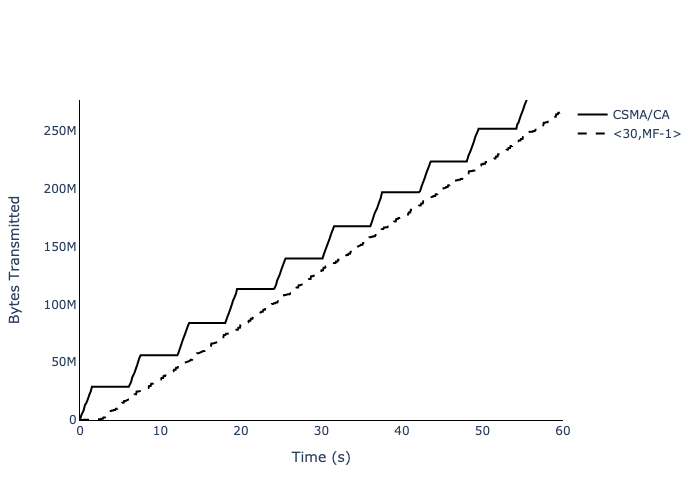}
\caption{Change in the cumulative transmitted bytes in a non scheduled and TWT scheduled scenario.}
\label{fig: Bytes vs Time}
\centering
\end{figure}

When there are sudden large spikes in the packet traffic on the network, causing the ingress buffers to be filled, resulting in increased latency experienced by the end users. This is usually experienced in the Wi-Fi scenarios as the Wi-Fi link is, generally, the bottle-neck link in the network. While QoS filters and EDCA do offer respite from this, it is still desirable to reduce such traffic ``burstiness'' by smoothening the incoming traffic. The graph in Figure~\ref{fig: Bytes vs Time} illustrates the smoothening of the bursts (3MB worth of data burst every 6 seconds) with TWT schedules. This smoothening could also allow the control of TCP congestion
window, hence reducing the overall packet flow into the ingress buffers throughput the network and subsequently packet latency, by setting appropriate TWT cycles on the final Wi-Fi link, a effect similar to server side rate control mechanism suggested by the authors in~\cite{b22}.

TWT scheduling can help control the burstiness of the traffic, allowing better performance for non-scheduled clients and the network as a whole. We also investigate the direct effect it has on latency. Toward this end, we inject bursty traffic from the traffic server to Clients 1, 2, and 5, while Client 6 acts a ping client whose latency we measure. The clients are then scheduled with the same $\left<25, MF-1\right>$ TWT schedules.
Table~\ref{Table: Change in ping} shows almost a 3 fold increase in the ping times when we introduce bursty traffic. We expect a improvement in ping times with TWT, but in reality, we consistently see worse ping latency for the TWT clients, a result similar to the latency reported by~\cite{b4}. We also observe ping times improving dramatically as we increase the MFs of the schedules, an effect that may be important considering scheduling of latency sensitive applications.

\begin{table}[!t]
    \scriptsize
    \centering
    \resizebox{\textwidth}{!}{%
    \begin{tabular}{|c|c|c|c|c|c|c|c|}
    \hline
    \multirow{2}{*}{Iteration No.} & \multicolumn{5}{c|}{Ping Latency}  \\ \cline{2-6} 
    &No Traffic& Bursty Traffic&Bursty Traffic with TWT&Bursty Traffic with TWT MF 5&Bursty Traffic with TWT MF 10\\ \hline
    1&95.387 ms/35.398 ms	&362.302 ms/289.117 ms	&482.069 ms/582.244 ms&78.499 ms/27.465 ms&68.966 ms/36.582 ms \\ \hline
    2&96.455 ms/31.937 ms	&389.091 ms/173.721 ms	&463.131 ms/592.834 ms&88.751 ms/39.365 ms&79.821 ms/55.005 ms \\ \hline
    3&99.550 ms/37.262 ms	&359.787 ms/151.994 ms	&502.783 ms/523.630 ms&85.197 ms/30.999 ms&68.701 ms/51.918 ms \\ \hline 
    4&114.479 ms/59.565 ms	&361.334 ms/161.747 ms	&491.259 ms/551.511 ms&80.819 ms/29.304 ms&49.123 ms/21.093 ms \\ \hline
    5&100.974 ms/29.679 ms	&378.431 ms/229.145 ms	&517.133 ms/561.596 ms&70.489 ms/39.761 ms&50.191 ms/28.376 ms\\ \hline																	
    \end{tabular}
    }
    \caption{The ping performance of TWT when bursty traffic is injected under different scenarios.}
    \label{Table: Change in ping}
    \end{table}

\section{Conclusion and Future Work}
\label{sec: Conclusions}
As CSMA/CA provides random channel access, it is difficult to provide performance guarantees in Wi-Fi networks with it, particularly in a standard compliant manner. In this work, we make use of a feature, Target Wake Time (TWT), in Wi-Fi 6 to provide such performance guarantees by deterministically scheduling the clients' channel access.

Such deterministic scheduling allows for a collision free operation of Wi-Fi in a congested scenario, thus allowing us to provide a minimum performance guarantee to every client.

In this work, we first set up an experimental testbed using commercially available off-the-shelf TWT and non-TWT-capable clients and other network components, and gain insights on TWT's operation in such a testbed. Using these insights, we formulate an optimization problem to determine the optimum deterministic schedule for the TWT-capable clients and values of various system parameters to maximize the system throughput. We then configure our testbed using these values and show that the corresponding deterministic schedule not only leads to a higher overall system throughput but also higher throughput for TWT-capable clients compared to the scenario where all clients use random channel access.

Much work remains to be done. We would like to repeat our experiments over a larger testbed and validate our findings. Further, extending this work to multi-AP scenarios with heterogeneous traffic is an interesting and practical direction. We have observed that in commercially available consumer-grade Wi-Fi 6 compliant clients, TWT firmware and its configurability varies across OEMs and implementations show peculiar behavior, such as accepting only some commands, unexpected teardown of TWT schedules, TWT being operational only when the device is not plugged into the charger, and variations in values of $WT_{max}$ etc. A better and uniform firmware support for TWT in off-the-shelf clients may allow us to reap much larger gains in network performance with TWT and it is a direction worth pursuing.

\end{document}